\documentstyle[12pt]{article}

\begin{document}
\title{Upper bound on the supersymmetry breaking scale in supersymmetric
$SU(5)$ model}
\author{N.V.Krasnikov \thanks{E-mail address: KRASNIKO@MS2.INR.AC.RU}
\\Institute for Nuclear Research\\
60-th October Anniversary Prospect 7a,\\ Moscow 117312, Russia}
\date{December,1994}
\maketitle
\begin{abstract}
The status of coupling constant unification in standard supersymmetric
$SU(5)$ model and its extensions is discussed. Taking into account
uncertainties related with the initial coupling constants and threshold
corrections at the low and high scales we find that in standard supersymmetric
$SU(5)$ model the scale of the supersymmetry breaking could be up to
$10^8$ Gev. In the extensions of standard $SU(5)$ model it is possible to
increase the supersymmetry breaking scale up to $10^{11}$ Gev.
\end{abstract}
\newpage

There has recently been renewed interest \cite{1}-\cite{12} in grand
unification business related with the recent LEP data which allow to measure
$\sin^{2}(\theta_w)$ with unprecendented accuracy. Namely, the world
averages with the LEP data mean that the standard nonsupersymmetric $SU(5)$
model \cite{13} is ruled out finally and forever (the fact that the
standard $SU(5)$ model is in conflict with experiment was well known
\cite{14,15} before the LEP data) but maybe the most striking and impressive
lesson from LEP is that the supersymmetric extension of the standard
$SU(5)$ model \cite{16}-\cite{18} predicts the Weinberg angle
$\theta_w$ in very good agreement with experiment. The remarkable success
of the supersymmetric $SU(5)$ model is considered by many physicists as
the first hint in favour of the existence of low energy broken supersymmetry
in nature. A natural question arises: is it possible to invent
nonsupersymmetric generalizations of the standard $SU(5)$ model
nonconfronting the experimental data or to increase the supersymmetry
breaking scale significantly. In the $SO(10)$ model the introduction of
the intermediate scale $M_I \sim 10^{11} Gev$ allows to obtain the Weinberg
angle $\theta_w$ in agreement with experiment \cite{19}. In refs.\cite{20,21}
it has been proposed to cure the problems of the standard $SU(5)$ model
by the introduction of the additional split multiplets $5 \oplus \overline{5}$
and $10 \oplus \overline{10}$
in the minimal $3(\overline{5} \oplus 10)$ of the $SU(5)$ model.
In ref.\cite{22} the extension of the standard $SU(5)$ model with
light scalar coloured octets and electroweak triplets has been proposed.

In this paper we discuss the coupling constant unification in standard
supersymmetric $SU(5)$ model and its extensions. Taking into account
uncertainties associated with the initial gauge coupling constants
and threshold corrections at the low and high scales we conclude
that in standard supersymmetric $SU(5)$ model the scale of the
supersymmetry breaking could be up to $10^8$ Gev. We find also that
in the extensions of the standard $SU(5)$ supersymmetric model it is
possible to increase the supersymmetry breaking scale up to $10^{11}$ Gev.

The standard supersymmetric $SU(5)$ model \cite{16}-\cite{18} contains three
light supermatter generations and two light superhiggs doublets.
A minimal choice of massive supermultiplets at the high scale is
$(\overline{3},2,\frac{5}{2}) \oplus c.c.$ massive vector supermultiplet
with the mass $M_v$, massive chiral supermultiplets $(8,1,0), (1,3,0), (1,1,0)$
with the masses $m_8, m_3, m_1$ (embeded in a 24 supermultiplet of $SU(5)$)
and a $(3,1,-\frac{1}{3}) \oplus (-3,1,\frac{1}{3})$ complex Higgs
supermultiplet with a mass $M_3$ embeded in $5 \oplus \overline{5}$ of $SU(5)$.
In low energy spectrum we have squark and slepton multiplets
$ (\tilde{u},\tilde{d})_{L}, \tilde{u}^c_L, \tilde{d}^c_L,
(\tilde{\nu},\tilde{e})_L, \tilde{e}^c_L $ plus the corresponding
squarks and sleptons of the second and third supergenerations.
Besides in the low energy spectrum we have $SU(3)$ octet of gluino
with a mass $m_{\tilde{g}}$, triplet of $SU(2)$ gaugino with a mass
$m_{\tilde{w}}$ and the photino with a mass $m_{\tilde{\gamma}}$.
For the energies between $M_z$ and $M_{GUT}$ we have effective
$SU(3) \otimes SU(2) \otimes U(1)$ gauge theory.
In one loop approximation the corresponding solutions of the renormalization
group equations are well known \cite{18}. In our paper instead of the
prediction of $\sin^{2}(\theta_w)$ following ref.\cite{6} we consider the
following one loop relations between the effective gauge coupling constants,
the mass of the vector massive supermultiplet $M_v$ and the mass of the
superhiggs triplet $M_3$:
\begin{equation}
A \equiv  2(\frac{1}{\alpha_{1}(m_{t})} - \frac{1}{\alpha_{3}(m_t)}) +
3(\frac{1}{\alpha_{1}(m_t)} - \frac{1}{\alpha_{2}(m_t)}) = \Delta_{A} ,
\end{equation}
\begin{equation}
B \equiv  2(\frac{1}{\alpha_{1}(m_t)} - \frac{1}{\alpha{3}(m_t)}) -
3(\frac{1}{\alpha_{1}(m_t)} - \frac{1}{\alpha_{2}(m_t)}) = \Delta_{B} ,
\end{equation}
where
\begin{equation}
\Delta_{A} = (\frac{1}{2\pi})(\delta_{1A} + \delta_{2A} + \delta_{3A}) ,
\end{equation}
\begin{equation}
\Delta_{B} = (\frac{1}{2\pi})(\delta_{1B} + \delta_{2B} + \delta_{3B}) ,
\end{equation}
\begin{equation}
\delta_{1A} = 44ln(\frac{M_v}{m_t}) - 4ln(\frac{M_v}{m_{\tilde{g}}}) -
4ln(\frac{M_v}{m_{\tilde{w}}}) ,
\end{equation}
\begin{equation}
\delta_{2A} = -12(ln(\frac{M_v}{m_8}) + ln(\frac{M_v}{m_3})) ,
\end{equation}
\begin{equation}
\delta_{3A} = 6ln(m_{(\tilde{u},\tilde{d})_L}) - 3ln(m_{\tilde{u}^c_L}) -
3ln(m_{\tilde{e}^c_L}) ,
\end{equation}
\begin{equation}
\delta_{1B} = 0.4ln(\frac{M_3}{m_h}) + 0.4ln(\frac{M_3}{m_H}) +
1.6ln(\frac{M_3}{m_{sh}}) ,
\end{equation}
\begin{equation}
\delta_{2B} = 4ln(\frac{m_{\tilde{g}}}{m_{\tilde{w}}}) +
6ln(\frac{m_8}{m_3}) ,
\end{equation}
\begin{equation}
5\delta_{3B} = -12ln(m_{(\tilde{u},\tilde{d})_L}) + 9ln(m_{\tilde{u}^c_L}) +
6ln(m_{\tilde{d}^c_L}) -
6ln(m_{(\tilde{\nu} ,\tilde{e})_L}) + 3ln(m_{\tilde{e}^c_L})
\end{equation}
Here $m_h$, $m_H$ and $m_{sh}$ are the masses of the first light Higgs
isodoublet, the second Higgs isodoublet and the isodoublet of superhiggses.
The relations (1-10) are very convenient since they allow to determine
separately two key parameters of the high energy spectrum of $SU(5)$ model,
the mass of the vector supermultiplet $M_v$ and the mass of the chiral
supertriplet $M_3$. Both the vector supermultiplet and the chiral supertriplet
are responsible for the proton decay in supersymmetric $SU(5)$ model \cite{18}.
In standard nonsupersymmetric $SU(5)$ model the proton lifetime
due to the massive vector exchange is determined by the formula \cite{23}
\begin{equation}
\Gamma(p \rightarrow e^{+} \pi^{o})^{-1} = 4 \cdot 10^{29 \pm 0.7}
(\frac{M_v}{2 \cdot10^{14} Gev})^{4} yr
\end{equation}
In supersymmetric $SU(5)$ model the GUT coupling constant is
$\alpha_{GUT} \approx \frac{1}{25}$ compared to
$\alpha_{GUT} \approx \frac{1}{41}$ in standard $SU(5)$ model, so we
have to multiply the expression (11) by factor $(\frac{25}{41})^2$.
{}From the current experimental limit \cite{24}
$\Gamma(p \rightarrow e^{+} \pi^{o})^{-1} \geq 9 \cdot 10^{32} yr $
we conclude that $M_v \geq 1.2 \cdot 10^{15} Gev$. The corresponding
experimental bound on the mass of the superhiggs triplet $M_3$ depends
on the masses of gaugino and squarks \cite{25,26}. In our calculations
we use the following values for the initial coupling constants \cite{24,27}:
\begin{equation}
\alpha_{3}(M_z) = 0.120 \pm 0.07 ,
\end{equation}
\begin{equation}
\sin^{2}_{\overline{MS}}(\theta_w)(M_z) = 0.2319 \pm 0.0005 ,
\end{equation}
\begin{equation}
(\alpha_{em,\overline{MS}}(M_z))^{-1} = 127.79 \pm 0.13
\end{equation}
For the top quark mass $m_t = 174 Gev$ after the solution of the corresponding
renormalization group equations in the region $ M_z \leq E \leq m_t$ we find
that
\begin{equation}
A = 184.45 \pm 0.68 \pm 0.92 ,
\end{equation}
\begin{equation}
B = 13.31 \pm 0.24 \pm 0.92
\end{equation}
Here the first error is the "electroweak" error and the second error is
the "strong coupling" error. An account of two loop corrections leads to
the appearance of the additional factors
\begin{equation}
\delta_{4A,4B} = 2(\theta_1 - \theta_3) \pm 3(\theta_1 -
\theta_3) ,
\end{equation}
\begin{equation}
\theta_{i} = \frac{1}{4\pi}\sum_{j=1}^{3} ln[\frac{\alpha_{j}(M_v)}
{\alpha_{j}(m_t)}]
\end{equation}
Here $b_{ij}$ are the two loop $\beta$ function coefficients.
Let us start from the expression (1) and assume that the masses of the
octet supermultiplet and the masses of the  triplet supermultiplet
coincide with the mass of the vector supermutiplet $M_v$. We shall neglect
the variation of the low energy spectrum (we assume that all the squarks
and the sleptons have the same mass). Numerically we find that
\begin{equation}
M_v = 2.0 \cdot 10^{16 \pm 0.05 \pm 0.07} Gev
\end{equation}
for the $M_{SUSY} \equiv (m_{\tilde{g}}m_{\tilde{w}})^{\frac{1}{2}} = 174 Gev$.
{}From the lower bound on the value of the mass of the vector bosons
responsible
for the baryon number nonconservation we find an upper bound on the
value of the supersymmetry breaking parameter $M_{SUSY} \leq 2\cdot 10^{8}
Gev$.
For $M_{SUSY} = 10^{8};10^{7};10^{6};10^{5};10^{4};10^{3};10^{2} Gev$
we find that $M_v = (1.4\cdot 10^{15};1.8\cdot 10^{15};3.0\cdot 10^{15};
5.1\cdot 10^{15};10^{16}; 1.6\cdot 10^{16};
2.4\cdot 10^{16})\times10^{\pm 0.05 \pm 0.07} Gev$.
The uncertainty in the masses of the coloured octet $m_8$ and electroweak
triplet $m_3$ leads to the uncertainty
$(\frac{m_{8}m_{3}}{M_{v}^2})^{\frac{1}{3}}$ for the
supersymmetry breaking scale $M_{SUSY}$. The uncertainty due to
the difference of squark and slepton masses is small for the realistic
spectrum when the difference in masses is less than 3 and we shall neglect
it.

Let us consider now the equation (2). For $M_{SUSY} = m_t$ and
$m_{\tilde{g}} = \frac{\alpha_{3}(m_t)}{\alpha_{2}(m_t)}m_{\tilde{w}}$
\cite{18} in the assumption that all squark and slepton masses coincide ,
the masses of the coloured octet $m_8$ and electroweak triplet $m_3$
are equal to the mass of the vector supermultiplet $M_v$
and the masses of superhiggses and Higgs isodoublets are equal to
$M_t$ we find that
\begin{equation}
M_3 = 6.6\cdot10^{14 \pm 0.27 \pm 1.05} Gev
\end{equation}
It should be noted that from the nonobservation of the proton decay
the bound on the mass of the Higgs triplet for $M_{SUSY} = m_t$ is
\cite{18} $M_3 \geq O(10^{16}) Gev$. When we increase the value of
$M_{SUSY}$ two scenario are possible. According to the first scenario
only the single Higgs isodoublet is light with a mass $O(M_z)$ and the
masses of the second Higgs isodoublet and superhiggses are of the order
of $M_{SUSY}$. In the second scenario the first Higgs isodoublet and
superhiggses are relatively light with the masses $O(M_z)$ (or superhiggses
are slightly heavier with the mass $O(1 Tev)$) and only the second Higgs
isodoublet is relatively heavy with the mass $O(M_{SUSY})$.
We have investigated two scenario. In our investigation we have used
an upper bound $M_3 \leq  3M_v$ \cite{5} which comes from the requirement
of the applicapability of the perturbation theory. Taking into account
uncertainties in the determination of the parameter $B$ we have found that
in the first scenario $M_{SUSY} \leq 10^{5} Gev $  and in the second scenario
$M_{SUSY} \leq 10^{8} Gev $. If we assume that the difference between
the masses of coloured octets and coloured triplets could be up to factor 3
then for $\frac{m_8}{m_3} =3$ we find that in the first scenario the
supersymmetry breaking scale could be up to $10^{7} Gev$. It should
be noted that the proton lifetime due to the exchange of the Higgs
supertriplet is proportional to $m^{2}_{sq}$ (here we assume that
$m_{sq} \sim M_{SUSY}$ ) and it does not contradict to the nonobservation of
the proton decay for big $M_{SUSY}$.

It is instructive to consider the supersymmetric $SU(5)$ model with
relatively light coloured octet and triplets. For instance, consider the
superpotential
\begin{equation}
W = \lambda \sigma(x)[Tr(\Phi^{2}(x)) -c^2] ,
\end{equation}
where $\sigma(x)$ is the $SU(5)$ singlet chiral superfield and $\Phi(x)$ is
chiral 24-plet in the adjoint representation. For the superpotential (21)
the coloured octet and electroweak triplet chiral superfields remain
massless after $SU(5)$ gauge symmetry breaking and they acquire the
masses $O(M_{SUSY})$ after the supersymmetry breaking. So in this scenario
we have additional relatively light fields. Lower bound on the mass
of the vector bosons leads to the upper bound on the supersummetry
breaking scale
$M_{SUSY} \leq 10^{11} Gev $
In order to satisfy the second equation for the mass of the Higgs triplets
let us introduce in the model two additional superhiggs 5-plets. If we assume
that after $SU(5)$ gauge symmetry breaking the corresponding Higgs triplets
acquire mass $O(M_v)$ , the light Higgs isodoublet has a
mass $O(M_z)$ , the second Higgs isodoublet and superhiggses have masses
$O(M_{SUSY})$ then we can satisfy the equation (2) for
$M_{SUSY} \sim 10^{11} Gev$. According two other scenario playing with
relatively light octets and triplets we can increase the grand unification
scale up to $O(10^{18}) Gev$ that is welcomed from the point of view of
the string unification scenario. In this case the supersymmetry breaking scale
is $M_{SUSY} \sim 10^{8} Gev$. For such value of the supersymmetry breaking
scale one can satisfy the equation (2) even without the introduction
of the additional superhiggs 5-plets. It is possible also to have grand
unification scale $M_v =  10^{17} Gev$ and $M_{SUSY} \leq 1 Tev$ if octet
and triplet are lighter than the vector supermultiplet by factor 100.

In conclusion let us formulate our main results. We have found that
in standard supersymmetric $SU(5)$ model with coloured octet
and triplet masses $O(M_v)$ the nonobservation of the proton decay
leads to the upper bound $M_{SUSY} \leq 2\cdot 10^{8} Gev$ on the
supersymmetry breaking scale. This bound does not contradict to the
equation for the superhiggs triplet mass in the second scenario.
In the first scenario it is possible to have the supersymmetry breaking
scale $M_{SUSY}$ up to $O(10^{7}) Gev$ provided that the difference
between the coloured octet and electroweak triplet masses
is $\frac{m_{\tilde{g}}}{m_{\tilde{w}}} = O(3) $ . For the case when the
octets and triplets have the masses $O(M_{SUSY})$ it is possible to
increase the supersymmetry breaking scale up to $O(10^{11}) Gev$ , however
in this case in order to satisfy the equation for the superhiggs triplet
mass we have to introduce additional relatively light pair of the superhiggs
doublets. It is possible also to increase the value of the grand unification
scale up to $O(10^{18}) Gev$ for the case when octets and triplets have
the masses $O(M_{SUSY})$ and $M_{SUSY} = O(10^{8}) Gev$ without the
introduction
of the additional light Higgs superdoublets. It should be noted that
estimate on the supersymmetry breaking scale in supersymmetric $SU(5)$ model
$ M_{SUSY} = 10^{3.0 \pm 0.8 \pm 0.4} Gev$ has been obtained in ref.\cite{28}
in the assumption that $M_v = M_3 =m_3 = m_8$. Our analysis demonstrates
that the value of $M_v$ depends rather weekly on the high and low energy
uncertainties in the determination of the spectrum and low energy effective
coupling constants. In the extraction of the bound on the value of
$M_{SUSY}$ our crusial assumption was the inequality \cite{5}
$ M_3 \leq 3 M_v$. The obtained bound on the $M_{SUSY}$ depends rather
strongly on the details of the high energy spectrum (on the
splitting between octet and triplet masses) and on the initial value of the
strong coupling constant. It should be noted that for $M_{SUSY} \geq O(1) Tev$
we have the fine tuning problem for the electroweak symmetry breaking scale.

I am indebted to the collaborators of the INR theoretical department
for discussions and critical comments. The research described in this
publication was made possible in part by Grant N6G000 from the International
Science Foundation and by Grant 94-02-04474-a of the Russian Scientific
Foundation.

\end{document}